\documentclass[sigconf]{acmart}

\usepackage{graphicx}
\usepackage{bibentry}
\usepackage{subcaption}

\AtBeginDocument{%
  \providecommand\BibTeX{{%
    \normalfont B\kern-0.5em{\scshape i\kern-0.25em b}\kern-0.8em\TeX}}}

\RequirePackage{color}\definecolor{RED}{rgb}{1,0,0}\definecolor{BLUE}{rgb}{0,0,1} 

\copyrightyear{2026}
\acmYear{2026}
\setcopyright{cc}
\setcctype{by}
\acmConference[CHI '26]{Proceedings of the 2026 CHI Conference on Human Factors in Computing Systems}{April 13--17, 2026}{Barcelona, Spain}
\acmBooktitle{Proceedings of the 2026 CHI Conference on Human Factors in Computing Systems (CHI '26), April 13--17, 2026, Barcelona, Spain}
\acmPrice{}
\acmDOI{10.1145/3772318.3791828}
\acmISBN{979-8-4007-2278-3/2026/04}

\begin{document}

\author{Barry Brown}
\orcid{0000-0002-9710-6607}
\email{barry@di.ku.dk}
\orcid{}
\affiliation{%
  \institution{University of Copenhagen}
  \city{Copenhagen}
  \country{Denmark}
  }
  
 \affiliation{%
  \institution{and Stockholm University}
   \city{Stockholm}
 \country{Sweden}
}

\author{Hannah Pelikan}
\orcid{0000-0003-0992-5176}
\email{hannah.pelikan@liu.se}
\orcid{}
\affiliation{%
  \institution{Department of Culture and Society, Linköping University}
  \city{Linköping}
  \country{Sweden}
  }

\author{Mathias Broth}
\orcid{0000-0002-9710-6607}
\email{mathias.broth@liu.se}
\orcid{}
\affiliation{%
\institution{Department of Culture and Society, Linköping University}  
\city{Linköping}
  \country{Sweden}
  }

\renewcommand{\shortauthors}{Brown, Pelikan, Broth}

\title{Walking with Robots: Video Analysis of Human-Robot Interactions in Transit Spaces}


\begin{abstract}
The proliferation of robots in public spaces necessitates a deeper understanding of how these robots can interact with those they share the space with. In this paper, we present findings from video analysis of publicly deployed cleaning robots in a transit space—a major commercial airport, using their navigational ``troubles'' as a tool to document what robots currently lack in interactional competence. We demonstrate that these robots, while technically proficient, can disrupt the social order of a space due to their inability to understand core aspects of human movement: mutual adjustment to others, the significance of understanding social groups, and the purpose of different locations. In discussion we argue for exploring a new design space of movement: socially-aware movement. By developing ``strong concepts'' that treat movement as an interactional and collaborative accomplishment, we can create systems that better integrate into the everyday rhythms of public life.
\end{abstract}



\maketitle

\section{Introduction}   
One of the most basic and important features for a publicly deployed robot is its ability to move in space safely. Interaction with others in that space—the problem of `pedestrian navigation'—has gathered considerable technical research as researchers have explored how to design a robot that does not collide with, or otherwise disrupt, others' movement \cite{due_walk_2023, pelikan_encountering_2024, mavrogiannis_core_2023}. While this can be approached as a technical problem, designing a system that can navigate successfully is also a `social' problem \cite{brown_trash_2024,erel_robots_2019}. That is to say, when humans move around in space, we manage our movements through our interactions with others through an understanding of others' motives, activities, relationships, etc. Spaces are arranged with different functions so, for example, there is a difference between waiting in front of or behind a counter, blocking a doorway, and so on. 

In this paper, we use data from publicly deployed airport cleaning robots to explore some of the `social troubles' that these robots encounter navigating a transit space. Our focus is not on critiquing these robots themselves, but rather to use these robots as tools to document design challenges, `what robots cannot yet do' with regard to moving in public. During a period of video-ethnographic fieldwork, we collected six hours of video recordings of airport cleaning robots, deployed in a major European commercial airport. We followed two different robot cleaning systems during the normal daytime operating hours of this airport. In collaboration with the airport, we collected data around these two cleaning systems, recording video of the robots cleaning floors and their interactions with staff and passengers waiting at the airport. The study of this specific dataset is informed by additional observational visits, and more broadly ongoing work studying human-robot interaction in delivery and driving situations. We analyzed our data from an ethnomethodological and conversation analytic (EMCA) perspective, focusing on how interaction evolves moment by moment~\cite{randall_ethnography_2001,Sidnell_Stivers_2012}.

Our results are presented in three sections. The first explores `Where are you going?', how others' trajectories influence the movement of the robots we studied, and their interactions with objects in the environment. The orientation of robots to others' trajectories is a problem of surprising complexity, in that—as with all multi-agent modeling \cite{yasuda_autonomous_2021}—we need to understand the extent to which adaptations to others' movements may result in them in turn modifying their own behaviour. In particular we draw on discussions of the `halting problem' \cite{Brown_Broth_Vinkhuyzen_2023} where robots make sudden stops causing others to react by halting suddenly. This can result in a cascade where others become `stuck' not knowing whether to go or wait for a robot to pass. In our second section, we address `who are you with?'. This focuses on the problem of social groups and the relevance of who is together with whom for navigation. This example explores the importance of groups in public settings. Lastly, in our third section we look at `what are you doing there?' - the social organization of space and how movement takes into account this organization of that space. Public spaces are spaces of activity where different things happen depending on what kind of space it is, such as for example a transit space. We want to draw attention to the point that space is not an undifferentiated flat area but imbued with a range of different purposes and meanings that are relevant for robot design. 

We build on these findings to contribute three arguments about the importance of social understandings of space for robot movement designers. First, contributing to the emerging body of work in robot navigation, we discuss how we might approach the problem of designing `socially aware movement', and the design question of what movement is in terms of an understanding of space and what is going on in that space by working on the case of the airport as a transit space. Second, contributing to discussions of how HCI understandings can inform robot design, we discuss the challenge of robots that might be `good enough' to avoid collisions but who are not quite good enough to fit smoothly and unproblematically into public space. Lastly, we present three strong concepts for designing robot motion, building bridges between the theoretical and empirical findings of the paper. 

\section{Prior work}
We start by reviewing existing studies of robot behaviour in public, and transit spaces in particular, then move on to review studies of pedestrian navigation and walking more broadly, and lastly, research into robot navigation. 

\subsection{Video analysis of robots in public}
As robots have become increasingly common in public spaces — delivering, cleaning or even policing that space–these interactions have been studied in CHI and adjacent venues such as HRI (Human-Robot Interaction) \cite{sabanovic_robots_2006}, building on the growing corpus of work studying self-driving cars and buses. 

A major strand of work is focused on robot deployments for research purposes. Several robot designs have been deployed by researchers to explore how people react to robots in public, for instance how people dispose trash \cite{brown_trash_2024} or how robots could creatively contribute to public spaces \cite{Hoggenmueller_Hespanhol_Tomitsch_2020}. Much of this work focuses on people who are not using robots, but who encounter them as bystanders, or \textit{incidentally co-present persons} \cite{RosenthalvonderPütten_Sirkin_Abrams_Platte_2020}. Early observations of robots in public remarked how such bystanders interfered with robot movement, such as when children discovered they could play with a robot's sensors \cite{Brščić_Kidokoro_Suehiro_Kanda_2015}. 

An emerging body of ethnographic work followed commercial robot deployments, primarily of food delivery robots in Europe \cite{pelikan_encountering_2024, Dobrosovestnova_Schwaninger_Weiss_2022}, the US \cite{Weinberg_Dwyer_Fox_Martelaro_2023}, and South Korea \cite{Cheon_Shin_2025}. These robots are portrayed as fully autonomous, needing no support from co-present humans to navigate in public spaces. Yet, a recurring observation is that other people need perform \textit{accommodation work}—coordinating with robots in fleeting encounters \cite{pelikan_encountering_2024}, clearing spaces for robots and adapting to their movement. 
For example, \citet{Weinberg_Dwyer_Fox_Martelaro_2023} observed a delivery robot pilot program in Pittsburgh, highlighting the many fleeting interactions that robots have along the way as they are moving in the city, and how studying emergent interactions is critical to understand how urban robots are integrated into a particular space. Capturing an extreme form of accommodation work by bystanders, \citet{Dobrosovestnova_Schwaninger_Weiss_2022} describe how passers-by helped robots stuck in the snow in the early phases of a delivery robot rollout in Tallinn, arguing for how the `kawaii' nature of these delivery robots encourages bystanders' support.

With respect to cleaning robots in particular, \citet{Babel_Kraus_Baumann_2022} studied `coexistence' between a large commercial cleaning robot and people at a train station. This study described a number of emergent interactions that occur in a public transit space, such as noticing and evading the robot. Recently, an experimental study comparing a large sweeping robot and a smaller cleaning robot found that the size and movement of the robot affected how people moved around them \cite{Raab_Miller_Zeng_Jansen_Baumann_Kraus_2025}.

While cars are not normally thought of as robots, self-driving cars face some of the same interactional problems. For example, based on video data from social media platforms, \citet{Brown_Broth_Vinkhuyzen_2023} analysed interactions around self-driving cars in real traffic, describing what they call `the halting problem', which has also been identified in robots \cite{pelikan_encountering_2024,Du_Brščić_Kanda_2025}. This problem concerns the motion of self-driving cars, in that they often stop—or `halt'—where it is not clear why they have halted. In these situations the problem is not only that a car halts on the road, but also that it becomes unclear if the car is yielding for another road user or not. They describe situations where self-driving cars appear to wait for other cars, but actually start to move just as that other car moves. This sort of motion can lead to a stand-off situation where it is not clear who is going to go, and the self-driving car's sporadic movements (and responses) result in cars becoming momentarily stuck, trying to decide who will go and who will yield. 

Looking to make robots better at navigating public spaces, Human-Robot Interaction researchers have started to investigate specific additional interaction modalities, such as movement \cite{Raab_Miller_Zeng_Jansen_Baumann_Kraus_2025, Moore_Currano_Strack_Sirkin_2019}, sounds \cite{Pelikan_Jung_2023} or projected lights \cite{Yu_Hoggenmüller_Tomitsch_2023}, as well as challenging current robot morphologies \cite{Yuetal2025}. Looking beyond opportunities for physical design of the robot, researchers have also mapped how robots can fit into public spaces, calculating their `\textit{robotability}' – how welcoming spaces are for robots \cite{Franchietal2025} and mapping activities and other characteristics of public spaces into design frameworks \cite{Pelikan_public-space_2025}. Our work contributes an HCI perspective to these efforts, drawing particularly on the long history of ethnomethodology and conversation analytic work on interaction with technology. 

\subsection{Walking as Social Practice}
If we turn to human experiences in public space, walking is a core activity. As such, walking in public has attracted a significant amount of attention in work on the organization of everyday life. Ervin Goffman was one of the first to take an interest in participants' walking and mobility practices as social practice, describing what he famously called ``vehicular units''. He noted how pedestrians may move as ``singles'' or as part of a ``with'' \cite{goffman_alienation_1967, jacobsen_interaction_2022}. Building on this work, studies of walking practices have further developed using video recordings as data. \citet{turner_notes_1974} document the methodical work of what they call production (walking in particular ways) and recognition (seeing others walking in particular ways). This work observed that pedestrians avoid walking through groups of pedestrians walking together, and describes a range of methods available to pedestrians for this accomplishment. Building on this they also discussed the concept of `membership categories' \cite{sacks_lectures_1995}. For instance, even if separated by a certain distance, two people may still be seen as together when they are recognizable as forming a parent-child pair. 

Lee and Watson \cite{lee_final_1993} made multiple observations concerning the visibility of walking practices and the categorisation of others in public spaces. Studying crowds in the London Underground and the Belleville market in Paris, they note how people organise themselves into `flow files', as people moving in the same direction in a crowd follow one another, how one joins a queue recognizably as the last to do so, and how different situations have their own `standard pace'. For example, the speed and distance separating members of a queue is less than in other mobile activities. More recent studies have continued empirical work on walking ``mobile formations'' \cite{mcilvenny_moving_2014}, for instance pedestrians crossing the road \cite{merlino_crossing_2019}. Walking speed and participants' relative positions have been described by \citet{lynch_when_2023} as constitutive of actions, activities and also interpersonal relations during an ``outing'' in the countryside. 

Pedestrians' spatial and mobile coordination has been analysed in terms of sequentiality and turn-taking, demonstrating how movement and walking practices are communicative in social interaction, e.g., to initiate focused interaction \cite{mondada_emergent_2009} in a step-by-step way or to close an interactional sequence by ``walking away'' \cite{broth_walking_2013}. \citet{mondada_collaboratively_2022}, in studying street encounters between outreach workers and passers-by, describe how co-presence can be transformed into more focused encounters through a process where pedestrians adjust to each other’s moves in a fine-grained way, either going along with an approach or refusing it. Human-robot walking interaction has been studied by \citet{due_walk_2023} who compared how a robot and a guide dog can support visually impaired persons, together forming specific mobile assemblages that enable their mobility as a ``with''.

Work in transport and urban design has explored topics such as pedestrian route choice \cite{tong_principles_2022}, and how the design of urban spaces (such as sidewalk materials and slopes) can influence accessibility \cite{martinez-chao_urban_2024}. Empirical work looking at pedestrian route choice has (using large scale GPS data, for example) highlighted pedestrians' preference for slightly longer paths with smoother sidewalks, better lighting and more appealing urban environments \cite{basu_what_2023, tong_principles_2022}. One influential approach has been space syntax, demonstrating that the urban street network structure itself shapes pedestrian flows and wayfinding, with measures of connectivity, integration, and visibility predicting route popularity and movement patterns \cite{hutchison_network_2005, sharmin_meta-analysis_2018}. 

Transit spaces, such as airports and train stations, have received particular interest in urban studies (e.g., \cite{lofland_world_1985}). These are places where people often only pass through on their way to somewhere else, but that nevertheless are regularly designed to cater for travellers' various biological needs (food, toilets, rest) and often also offer spaces for work. Marc Augé describes them as \textit{non-places} in contrast to anthropological spaces \cite{howe_non-places:_1995}, a notion that has been challenged as research has mapped the connection between place (making) and mobility. Airports in particular have been the subject of such place-making efforts, with literary works taking their fluctuating and rhythmic nature as a source of inspiration for novels, such as the writers in residency at Heathrow airport \cite{Eldelin_Nyblom_2021}.

\subsection{Robot Navigation}
Making a robot move unaccompanied in crowded public spaces is a significant technical challenge, made possible by large research and industry investments into motion planning and navigation. Most relevant to this work is the body of work on \textit{social robot navigation}, which aims to translate insights about human ways of moving and using space into robot design \cite{Francis-socialnavigation2025}. This work develops technically how robots can participate in social interaction. In a recent review paper, \citet{mavrogiannis_core_2023} summarise the challenges of social robot navigation as one of \textit{proxemics}, the distance that robots should keep to humans, \textit{formations}, how people move together in a group and \textit{intentions}, where someone will move next. Summarising the discussions from a social navigation symposium, \citet{Francis-socialnavigation2025} identify similar challenges: To navigate socially, robots have to maintain a safe and comfortable distance. They should move around people in an appropriate way, moving in legible ways, and adhering to politeness and other social norms. Finally, robots need to predict and proactively accommodate the behaviour of others. These authors identify the specific context as important, highlighting cultural, diversity and environmental elements that robots should be designed for.

Teaching robots to keep not only a safe, but also comfortable distance is inspired in part by Edward Hall's work on proxemics \cite{hall1973silent}, which described the distance that people keep to each other depending on their relationship. The technical work on proxemics typically centres on questions about what  the right distance is that a robot should keep to a person \cite{Obaid_Sandoval_Zlotowski_Moltchanova_Basedow_Bartneck_2016, MummMutlu2011}. A challenge for such work is that there is no universal distance that is ``right'', but depending on the situation, culture or personality traits, the preferred distance to an approaching robot might vary \cite{Shen_Tennent_Claure_Jung_2018,Takayama_Pantofaru_2009}. Mobile settings where people and robots briefly pass each other, such as getting close when passing in narrow passages, further complicate this matter, since it may be acceptable to get close on one side but not on the other \cite{Kirbyetal2009}. 
Maintaining an adequate distance may not always be possible in crowded areas, leading robots to freeze \citep{Trautman_Krause2010}, coming to a complete halt when unable to predict trajectories without getting too close to people. Often braking happens quite harshly \cite{Du_Brščić_Kanda_2025} potentially causing confusion for those nearby.

Beyond stopping before hitting an obstacle, robots in public settings need to be able to recognise group formations, and people standing or walking in close proximity, in order to minimally adhere to social norms of movement. Research has been inspired by what \citet{Kendon_1990} described as facing formations, observing how people may arrange themselves spatially in different types of configurations. In early work, HRI researchers explored how robots could approach stationary groups, discovering for instance that it is preferred if robots approach people from the side rather than from the front \cite{Dautenhahn_Walters_Woods_Koay_Nehaniv_Sisbot_Alami_Siméon_2006}. Detecting whether people are part of a stationary group requires tracking the lower body \cite{Vazquezetal2015}, and recognising people who are standing in line requires modelling both orientation and distance to each other \cite{Nakauchi_Simmons_2002}. Recent work predicts membership even in mobile groups of pedestrians, based on people's orientation and distance to each other \cite{pmlr-v164-wang22e}.

To allow for mutual coordination of movement, robots need to indicate where they intend to move next \cite{Pascher_Gruenefeld_Schneegass_Gerken_2023}. Robot motion can be deliberately shaped to be more legible, helping humans to anticipate and coordinate with the robot more easily \cite{Dragan_Lee_Srinivasa_2013}. Designing motion to be expressive rather than just efficiently reaching a goal may involve moving on a curved trajectory rather than a straight line \cite{szafir_communicating_2015}, making it possible to recognise where a robot is heading. 

While research efforts in social robot navigation have taken inspiration from how humans navigate spaces in order to design robots, the majority of this work happens in simulations and lab experiments. In contrast, our study here explores the challenging environment of commercial cleaning robots moving in the  dense and dynamic setting of the airport.  Video analysis contributes to this body of work by providing an empirical grounding, connecting longstanding technical challenges with situated analyses of how people actually manage movement together. We aim to show not only that these problems persist outside the lab, but also that robots can be built on a richer understanding of movement as a social and sequential accomplishment. Our goal is to formulate design challenges grounded in ethnographic findings for the technical development of robots that can navigate public spaces; to establish a \textit{design space} for robotic motion.

\section{Method}
We use video data from a deployment of robots in a transit space to explore some of the `troubles' of robots navigating in public spaces, and what this exposes about the social nature of space. We should emphasise that we are not attempting to evaluate these robots in any sort of direct way. Instead we wish to understand what `robot movement in public settings' is, as a human computer interaction research topic, and in particular what challenges we can find for designing these systems. 

As researchers with an ethnomethodological perspective on activity \cite{garfinkel_studies_1967,heath_technology_2000}, we documented the `seen but unnoticed' aspects of robot interaction. As in related work in HCI, these findings often have interesting implications for design \cite{heath_technology_2000}. Our approach in particular was informed by the longstanding tradition of work within HCI and CSCW that uses video to look closely at the moment-by-moment interaction with technology \cite{brown_iphone_2013,flor_case_2010,heath_video_2010}. We also draw heavily on the long history of conversation analysis as an approach to understanding human interaction \cite{Sidnell_Stivers_2012,Hutchby_Wooffitt_2008}. This work has pioneered looking at interaction in terms of sequences of action, with an intense focus on small sections of data in an attempt to provide a detailed and deep rather than `broad' or summary analysis of the phenomena. These approaches have previously been applied in HCI to study, for example, conversational user interfaces and social robots \cite{Yamazaki_Yamazaki_Okada_Kuno_Kobayashi_Hoshi_Pitsch_Luff_vomLehn_Heath_2009,porcheron_voice_2018,Pelikan_Broth_2016}. 

\subsection{Data Collection: Video-Ethnographic Fieldwork}
For our data set we made video recordings of two different robot cleaning systems operating in a large European commercial airport. Being a transit space for passengers, the airport can be considered a ``perspicuous setting'' \cite{garfinkel_ethnomethodologys_2002} for documenting one-off encounters between particular machines and pedestrians, where the latter must thus draw on their ordinary competence and general expectations—rather than specific knowledge about a certain machine or system–in managing the encounter. 

The authors regularly travel through the airport we studied and are well familiar with its rhythms, observing the cleaning robots frequently while travelling. As a traveller however, it is difficult to know when and where robots will be deployed. To conduct more systematic video ethnographic fieldwork following the robots, we therefore established a collaboration with the airport cleaning staff who were responsible for the robot. 
Our video-recordings are therefore guided to some extent by the perspective of the cleaners who deployed the robots to clean specific areas. 

The robots are deployed on a daily basis to mop the floor of the airport, operated by the staff who also clean alongside the robots, sometimes re-cleaning areas where the robot had already passed. While the robots mop large areas, the human cleaners sweep the areas that the robot cannot access, such as underneath benches. While neither of these robots are customer service robots specifically, they do interact with different users (such as the cleaners who initiate the autonomous cleaning). However, our main focus was not this direct user interface, but the incidental interface that came from their interactions with others inhabiting these spaces. Earlier work on human-robot interactions has characterised this as `bystander' relations \cite{pelikan_encountering_2024}.

In collecting data, our broad approach was to study as `natural' as possible interactions between the robot and the people passing by it on a regular day at the airport. We therefore accompanied the robots during their normal daily tasks, in this case sweeping and mopping the `airside' floors of the airport. The first and second author recorded the robots with two handheld action cameras. 
The robots were set to clean in different areas and sections of the airport, which afforded different ways of recording. When cleaning a long hallway that was connecting different parts of the airport, we had to walk close to the robot to observe and videotape what was happening. When cleaning in a transit space where we could see the entire scene, we could remain stationary for longer stretches of time and did not have to walk among travellers as much. During one cleaning session, we placed a third 360° camera on the larger robot, which enabled us to stay a bit further away from the robot. 

Following conversation analytic data collection practice, we embrace that we can never fully remove ourselves from the scene but instead focus on the natural organisation of interaction, analysing our own involvement as observers when it is made relevant by the people in the scene \cite{Mondada_2012, Lynch_2002}. While some people acknowledged us with a brief glance and may have recognized us as a group of three–the cleaning staff with a tag and two people holding cameras–we were never asked directly who we were and what we were doing, but rather treated as part of the inventory of the airport.

\subsection{Data}
In collaboration with this airport, we collected six hours of video data of two different systems cleaning in three different public areas of the airport. The two robots were a larger, chest-height Nilfisk Liberty~SC60 and a smaller, hip-height Ecobot~40 cleaning robot. We recorded video of the robot cleaning floors and its interactions with staff and passengers. While we focused on collecting video data of the movement of the robots, we also interviewed the head of the cleaning unit, and recorded an in-situ interview with two members of the cleaning staff, who were in charge of managing and maintaining the robot. During our fieldwork we also observed a call to the robot manufacturer when one robot was not functioning as desired. \autoref{tab:data} provides an overview of the data we collected. In our analysis, we focus on the instances when the robot was cleaning autonomously.

\begin{table*}[]
    \centering
    \begin{tabular}{c|c|c|c}
         \textbf{Location} & \textbf{Time} & \textbf{Robot} & \textbf{Mode}\\
         \hline
         Terminal 3, F-gates storage space & early morning & Nilfisk Liberty SC60 & manual ride to cleaning location \\
         Terminal 3, E-gates hallway & early morning & Nilfisk Liberty SC60 & autonomous mopping (four rounds) \\
         Terminal 3, F-gate floors & morning & Nilfisk Liberty SC60 & manual ride in elevator \\
         Terminal 3, transfer hall & morning & Nilfisk Liberty SC60 & autonomous mopping \\
         Terminal 3, transfer hall & morning & Nilfisk Liberty SC60 & manual takeover\\
         Terminal 3, transfer hall & late morning & Nilfisk Liberty SC60 & autonomous mopping \\
         Terminal 3, F-gate storage space & late morning & Nilfisk Liberty SC60 & manual ride to storage location \\ 
         Terminal 2, B-gate storage space & lunch time & Ecobot 40 & manual drive to cleaning location \\ 
         Terminal 2, B-gate rotunda & lunch time & Ecobot 40 & customer support call\\
         Terminal 2, B-gate rotunda & lunch time & Ecobot 40 & autonomous sweeping\\
        Terminal 2, B-gate rotunda & lunch time & Ecobot 40 & manual takeover\\
        Terminal 2, B-gate storage space & afternoon & Ecobot 40 & manual drive to storage location \\

    \end{tabular}
    \caption{Overview of the collected data.}
    \label{tab:data}
\end{table*}

\subsection{Video Analysis}
Following conversation analytic practice \cite{Sidnell_Stivers_2012, Hutchby_Wooffitt_2008}, after discussing ethnographic observations in the author team and syncing the different camera perspectives, we went through all video material to build a collection of clips for further in-depth analysis. We did this by dividing the videos recorded in different areas of the airport amongst the authors who watched each video in full length to extract incidents for further joint analysis, focusing on cases of interaction between the robots and others in the public environment. 

We selected forty interaction `highlights'—looking for cases where the interaction seemed noteworthy—either because something went wrong, or an interaction was particularly smooth, or cases that seemed particularly unusual or typical \cite{heath_video_2010}. The length of these extracted videos varied between 10 seconds and 3 minutes. As is typical in an interactional approach to video, the goal is not to produce a reproducible taxonomy or overview of the data, but rather a close analysis of interactions, alongside a scenic understanding of how each phenomena fits into the scope of broader actions \cite{Schegloff_1993}. Methodologically, this involves detailed (often group) analysis of video extracts and their ongoing comparison within the corpus to achieve analytic rigour and validity. For this paper, we analysed the 40 fragments in depth in a series of group data sessions with the whole author team \cite{heath_video_2010}. Each extract was thus looked at as an individual, unique incident - but also inspected for exemplifying patterns that we could extrapolate to understand robot interaction more broadly.

Ultimately, we selected 22 shorter clips for further moment-by-moment analysis through transcription. These clips were between 10 seconds and 1 minute long and illustrate the problems the cleaning robots encounter in their interactions in the airport as well as how they do move unproblematically for much of the time. We first analysed these clips individually, creating transcripts and visualisations of the interaction in each clip. We then met with the whole author group to jointly analyse and contrast the clips, eventually making a selection of which examples to present in this paper. The clips showcase both robots and all three locations where the robots cleaned autonomously. They are representative of the interactions that we observed in the airport and of the moments we had selected in the broader dataset. We deliberately selected clips that bring out some of the more distinctive aspects of robot interaction in the clearest way. The clips do not have a formal relationship to the larger corpus in the sense that we aim to cover all types of behaviour as in qualitative human factors analyses as presented for example by \citet{Babel_Kraus_Baumann_2022}. Instead, since our aim was to develop findings that are generative for design \cite{Lupetti_Zaga_Cila_2021, hook_strong_2012}, we selected clips that illustrate phenomena that are particularly relevant for (and currently underexplored in) social robot navigation. 

\subsection{EMCA Approach}
We build on the emergent body of work that has used video to understand the complexity of human-robot interaction as real-time sequential phenomena \cite{brown_trash_2024}. Ethnomethodological and conversation analytic approaches (EMCA) have risen in prominence for understanding human-robot interaction \cite{pelikan_encountering_2024,Yamazaki_Yamazaki_Okada_Kuno_Kobayashi_Hoshi_Pitsch_Luff_vomLehn_Heath_2009}, and one strength of this approach is that it focuses on action, providing a toolkit of concepts for understanding the moment-by-moment unfolding nature of interaction, allowing analysts to grasp why the sequence and timing of movements can be important, and how sometimes a movement can have a different meaning if it happens even a second later (as in the case of a movement following, rather than pre-empting, another's move). We emphasise that this approach builds on the longstanding use of interaction video analysis as a method in HCI \cite{jordan1995interaction, button_technomethodology:_1996}. Our goal was not a quantitative or summative evaluation, criteria which are irrelevant to our analytic approach \cite{Schegloff_1993}.

For scholars in this tradition, public space has also been a rich domain in which a number of concepts and arguments have been developed. Notable, for example, is work on uninvited interactions in public \cite{llewellyn_streetwise_2008, Mondada_2022_trajectories}, Liberman's work on public space \cite{liberman_study_2019}, and  Laurier's studies of cafe and street interaction \cite{laurier_civility_2018}. EMCA has also developed a longstanding interest in sites of interactional trouble: moments of friction, outliers, or edge cases which while often not the main analytic focus, actually work to help us understand the social order. In HCI, notable work in this tradition would be Brown et al's work on the `troubles' that develop around self driving cars \cite{brown2017trouble}, and Reeves's work on performances in public space \cite{reeves_designing_2005}.

\subsection{Ethics and consent}
Our university, and the country in which the recordings took place, does not require institutional IRB approval for non-biomedical research. As such, following our national and institutional research regulations this research was conducted with an approach that relies on researchers' own explicit self-review, along with adherence to standardised (ACM) ethical guides and national data protection laws. Along with registering our data collection, as a team we conducted periodic discussions amongst ourselves to check and maintain our ethical approach. This included, in particular, protection of the data that was recorded, and consent from those who were directly involved. The recordings were made in what we would characterise as a public, or semi-public, place. The guidelines that we have followed do not require consent for recording in public places, or spaces where there is no `usual expectation of privacy'. Indeed, the airport authority that granted us permission to record itself conduct extensive video surveillance of this space. 

Practically, while recording we did not keep the camera on bystanders if they seemed uncomfortable, and following our interest in the actions around the robot, we only filmed those who were interacting with the robot in some way. While recording, we wore signs showing a video camera symbol and text explaining that we were ``recording for research'' (with a university logo). We walked close to the robot, with authorised personnel present, and the robot itself also had a similar sign on it alongside a 360 recording camera. During our recording we did not at any time hide that we were recording, our cameras were always held up visible to those around. We did not have anyone approach us to discuss the recordings or ask to withdraw. As it is still important to preserve anonymity and the privacy of those involved, we have blurred the faces in our video extracts here. 

Our analytic interest in how people spontaneously encounter and move around cleaning robots that are part of established work routines at the airport would have made approaching people for consent very difficult, as there are many hundreds of passers-by from our recordings. In contrast to studies in public where the research team deploys a novel technology, our aim was to document how the airport's own cleaning robots move among people. As discussed in previous CHI studies conducted in public spaces (such as \cite{brown_trash_2024,pelikan_people_2025}), asking consent from everyone passing through a public space is not practically feasible. We did however receive written permission from the authority that manages the airport and consent from the cleaning staff that guided us through the airport.

\section{Findings}
We present our findings in four sections. Each one of these sections will introduce a short video clip from our fieldwork, discuss some of our analysis of the interaction, and conclude with observations about the robots' interactions. Our four extracts (1) introduce \textit{the airport} as a transit space, explore (2) \textit{where you are going?}, (3) \textit{who are you with?} and (4) \textit{what is going on here?} 

\subsection{The airport as a transit space}
\begin{figure*}[h]
    \centering
    \includegraphics[width=\linewidth]{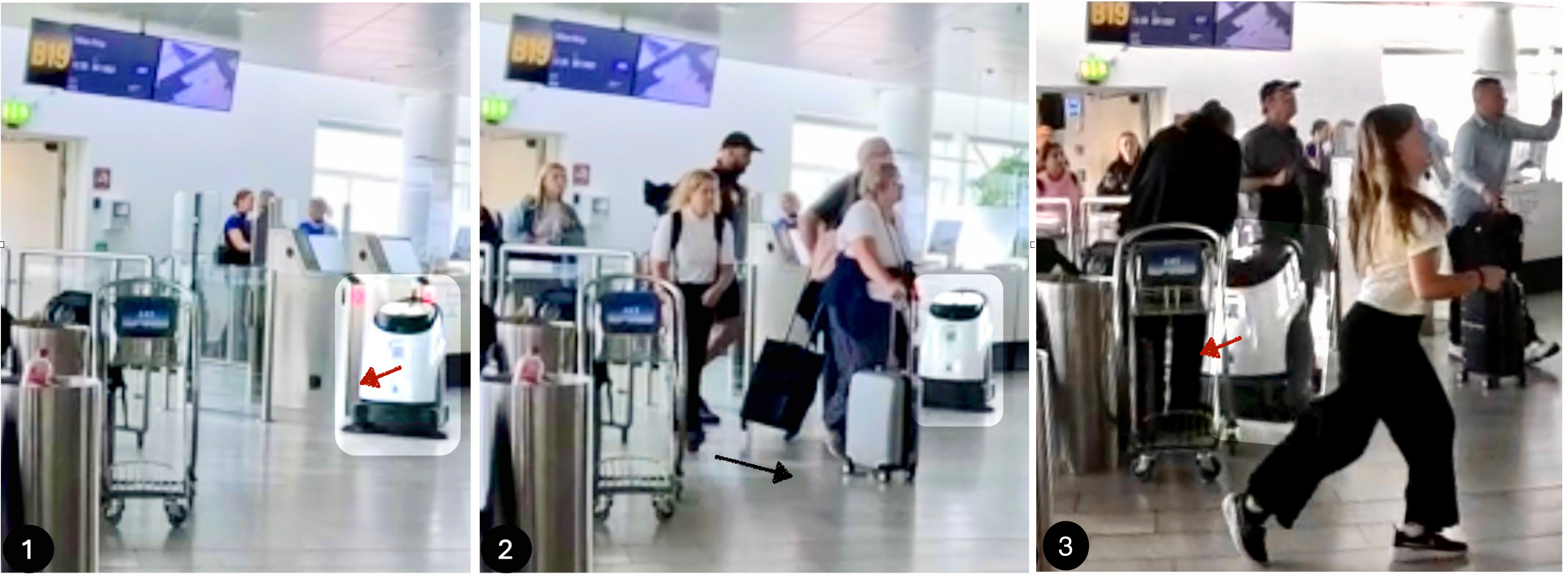}
    \caption{Robot cleaning at an airport gate. (1) moving to the gate when it is closed. (2) A few moments later, people disembark through that gate. (3) Robot gets stuck in the flow of people. As the crowd go around the robot it stops and although it does not block the way, arriving passengers need to walk around the robot which is positioned in front of the gate area.}
    \label{fig:gate}
    \Description{}
\end{figure*}

We start our findings with Figure \ref{fig:gate}. This shows a sequence from a clip of an Ecobot cleaning robot sweeping the area in front of an airport gate. At the start of our recording, the gate is not open and the robot manages to move around without causing any disturbances, cleaning the floor in the area around the gate and the gate counter (\autoref{fig:gate}.1).  
A few moments later, an arriving plane disembarks and passengers start to flow through the gate (\autoref{fig:gate}.2). The robot moves into the gate area and now becomes stuck within the crowd of disembarking passengers, stuck behind a luggage trolley, but also blocking the way for the passengers coming off the plane (\autoref{fig:gate}.3). Eventually the cleaning supervisor had to manually intervene and drive the robot away from the gate area.

When a robot is deployed to a public space it encounters a complex situated environment, where the appropriate behaviour for a location can change quickly from moment to moment. In particular, with airports as transit spaces, the arrival and departure of planes leads to dynamic rhythms of crowds of people appearing and disappearing. A space that was empty at one moment may suddenly be crowded by hundreds of people who are exiting a plane. This clip is emblematic of the challenges that robots face in moving through an airport. Not only do robots need to be able to understand human movement in space, and thread their own presence around that movement, but they also need to be aware of the complex \textit{social} nature of space. Pelikan et al. have recently drawn attention to the activity rhythms that are tied to public places and the ways in which they impact robots \cite{Pelikan_public-space_2025}. Transit spaces in particular are characterised by rhythms of large crowds of people that quickly appear and disappear. At one moment a space might be quiet and easily cleanable, in another it suddenly transforms into a lane through which hundreds of disembarking passengers must pass. 

The cleaning robots can move around the airport and clean in different spaces without physically colliding with others. Yet, can a robot understand the complexities of a social space to be able to \textit{smoothly} interact in that social space? How much does a robot need to understand about the nature of transit spaces as spaces with rapid shifts between anonymous and dull and buzzingly active \cite{howe_non-places:_1995, Eldelin_Nyblom_2021}? 
That is to say, beyond avoiding collisions, what does a robot need to understand to be able to successfully interact with others who are in that space without causing confusion and disruption? Currently, while the robots drive autonomously, it is the human cleaners who manage the social aspects of deploying the robots, manually taking over when robots get too close to people or preferably deploying robots at night when gates tend to be less busy. 

\subsection{``Where are you going?'': The halting problem}
\begin{figure*}[h]
    \centering
    \includegraphics[width=\linewidth]{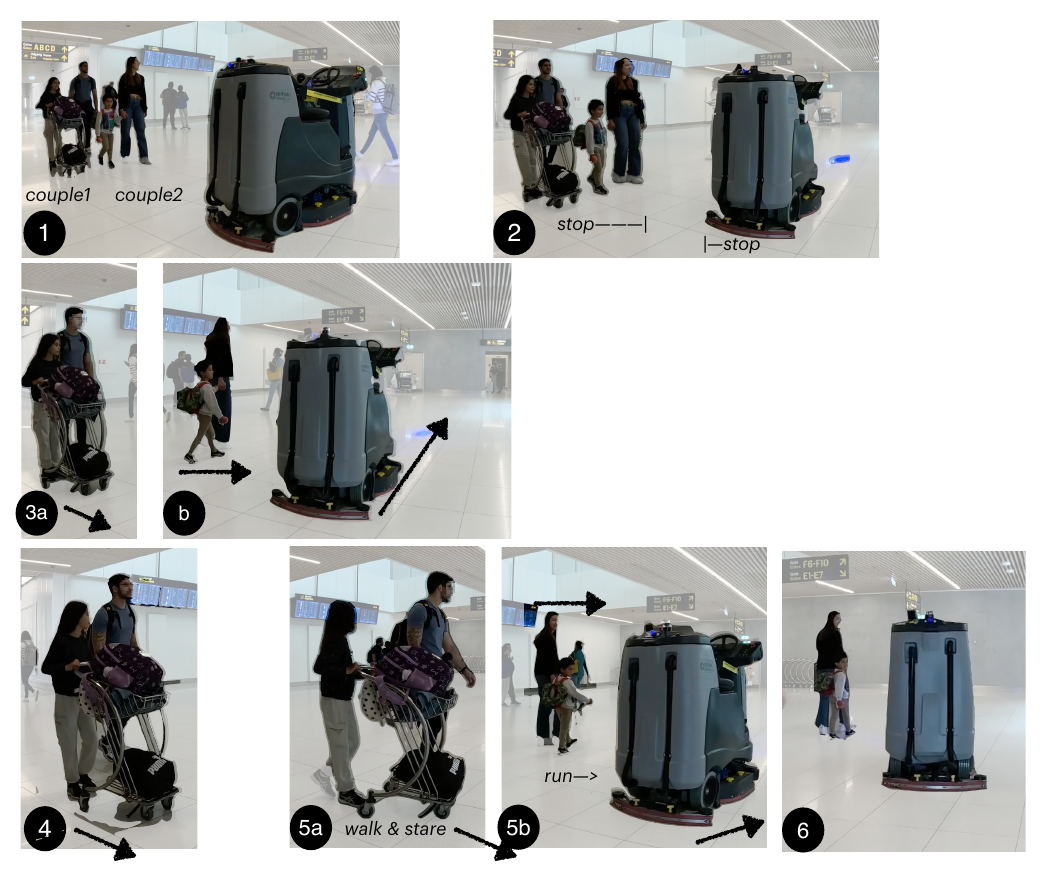}
    
    \caption{(1) Two separate groups (couple1 and couple2) encounter a cleaning robot in their path. (2) The robot stops, prompting both couples to stop. They now need to go round the robot as their path is blocked. The groups go in different directions (3a \& 3b), the first couple (3a) approach the robot which then starts to move (3b) as the second group head off to the right. The first couple continue (4) and then turn to look at the robot (5a) The robot starts to move rapidly (5b) and the second group need to move quickly to not be blocked by the robot. The mother starts to pull slightly (6) on her child's hand. Both groups stare at the robot as it first halted, and secondly as it starts to move and seemingly `race' the mother and child (5a \& 6).}
\label{fig:race}
\Description{}
\end{figure*}

In Figure \ref{fig:race}, two groups of passengers (a couple, and a mother and child) are walking across the hall where they meet a cleaning robot. While these two groups are physically close, they are not travelling together and go off in different directions later in the extract.  While the robot could continue moving (and then would likely be out of the way by the time the pedestrians got there), it instead halts—directly in the middle of the pedestrians' path. 

This causes the two groups to themselves stop with a small `bump' in their motion so they do not walk directly into the robot (\autoref{fig:race}.2). Here we have an emerging stand-off. The robot has `halted' and it is not clear what the pedestrians should do - should they wait for the robot or go themselves? There is a pause of a few seconds during which the robot does not move or show any intention of moving. The pedestrians then move themselves, walking around the robot, with one group moving to the left of the robot, and the other to the right. Very briefly afterwards the robot starts to move again  (\autoref{fig:race}.3). This now means that the robot is moving into the new path of the mother and child. The mother and child then have to `race' to overtake the robot (\autoref{fig:race}.5). The woman even has to pull on her child to get them to almost run, so they can get around the robot in time. Even then, the robot actually turns slightly to the left, seemingly trying to intercept them as they walk away. As the robot moves it is stared at by the first couple.

These sort of problems have previously been discussed as the `halting problem' \cite{Brown_Broth_Vinkhuyzen_2023}, and a version of this can be seen in \autoref{fig:race} with our airport robot. The halting problem points to how a stopped robot can lead to confusions for those around about a robot's intent.  Is it yielding—allowing another to go before it, or is it simply stuck? In this case the robot halts with little explanation—it is not clear to the pedestrians why it is halting (or for whom).  Is the robot going \textit{or} yielding? As they start to move the robot itself then also starts to move again—leading to the passengers having to hurry to outmanoeuvre the robot. The robot's stop was actually \textit{not} a `yield', causing some confusion and amusement for the travellers. 

As Brown \textit{et al} argue, a broader issue with robotic movement is when robots fail to make ongoing mutual adjustments to others' motions. Humans pre-emptively adjust their direction and speed to fit with the motions of others as their movements become visible. This can be particularly important in dense transit places where humans must squeeze around and between each other. Pedestrians continously monitor and mutually adjust so as to maintain space between each other and the environment. If a robot moves in a way that will conflict with others, this results in avoidance behaviour by pedestrians—the ongoing mutual adjustment of space. Yet this avoidance behaviour is not just a single individual action—it is also a signal to others that \textit{they themselves} should make adjustments in turn. So while the pedestrians make resultant avoidant behaviour in the case above, the robot's next action is actually the opposite of what it should do—it not only fails to take into account their new trajectory, but it also sets off in a direction that directly conflicts with that movement.

More broadly, we can see that human actions in space are `accountable' to others \cite{garfinkel_studies_1967} as sensible and projectable actions. That is to say they are readable and understandable in terms of where we are going.  This allows others to make subtle adjustments and create a smooth flow in sometimes congested spaces. But in this case the robot actually moved in ways that blocked this understanding—stopping and starting in seemingly chaotic ways. For these reasons the accountability of the robot failed \cite{laurier_what_2012}, leading to ongoing confusions for those who shared the space with the robot.

\subsection{``Who are you with?'': Understanding groups}
\begin{figure*}[h]
\centering
    \includegraphics[width=\linewidth]{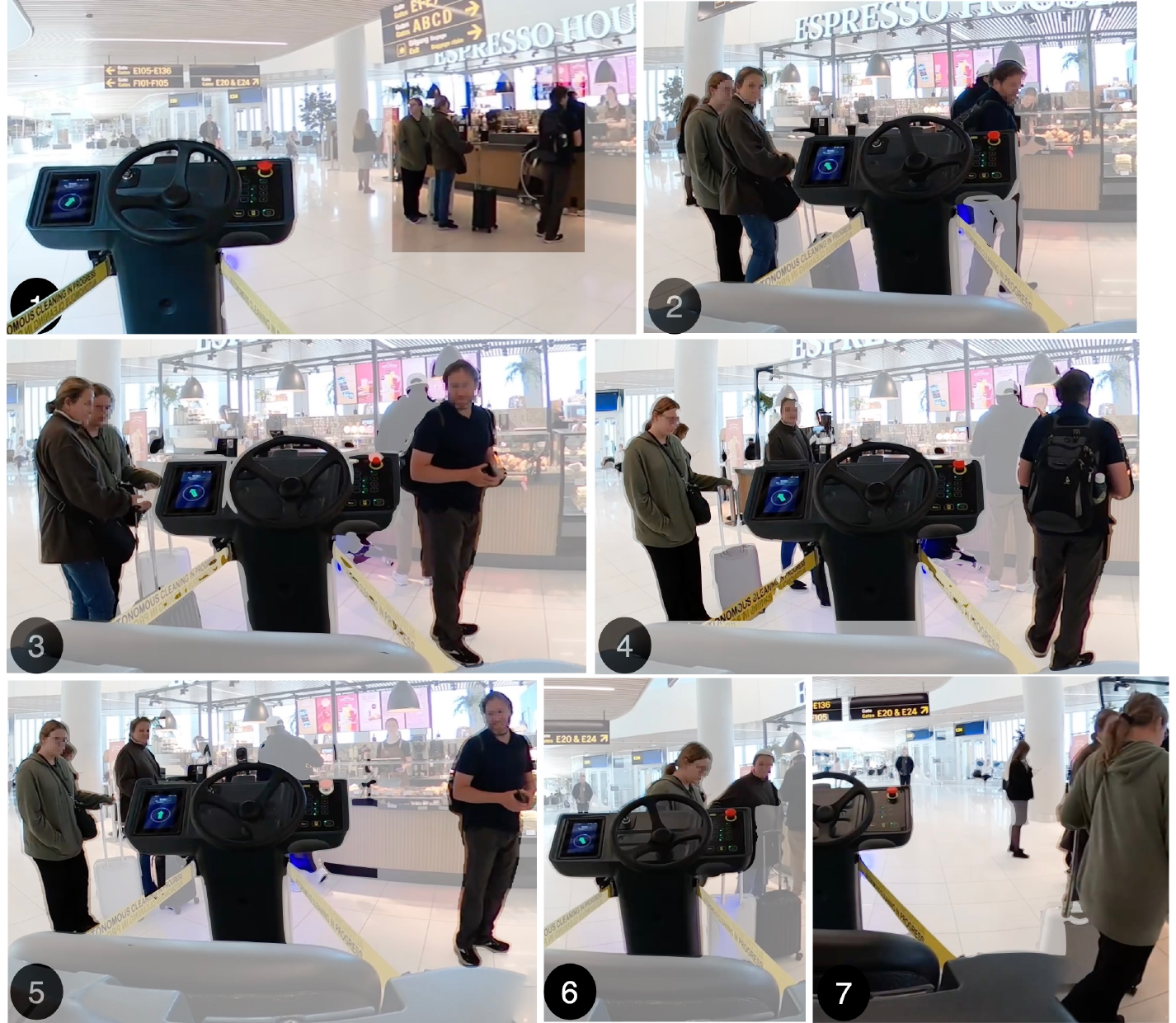}
\caption{A group are queuing for service at an airport cafe (1). The robot attempts to clean between the first and the next queue member (2). Despite attempts to move out of the way and accommodate the robot (3 \& 4) it stays motionless (5). Then it turns to move directly at one of the queue members (6), causing amusement amongst those waiting. Finally, it suddenly turns and goes around the queue on the left (7).}
\label{fig:queue}
\Description{}
\end{figure*}

For our next extract in Figure \ref{fig:queue}, we take the view from the robot itself, with a recording from a camera that we had mounted on top of the robot. As the extract starts, we see a group arranged in front of the cafe—three people are in a queue together waiting for service (\autoref{fig:queue}.1). The robot turns and heads towards the cafe service area (and the people queuing), directing itself towards a small gap between the first and the second member of the queue (\autoref{fig:queue}.2). The queue members move to accommodate the robot's motion—they separate around a potential path for the robot, with the man at the head of the queue taking a step forward and the two women behind him moving back. At this point the robot stops, wiggling slightly to the left and right. This attracts all queue members to stare at the robot, perhaps unclear as to its intentions (\autoref{fig:queue}.3). After a second or two of waiting, in which the robot does not move, the second queue member rearranges herself around the robot to make an even larger space for the robot to come forward (\autoref{fig:queue}.4). The robot's steering wheel then spins all the way to the right, all the way to the left (\autoref{fig:queue}.5), and then finally the robot does a turn to the left (\autoref{fig:queue}.6) and goes around the queue members on their far side. As it does so, the third queue member has to move out of the way to allow the robot to pass (\autoref{fig:queue}.7).

Continuing from our earlier example, the motion of the robot here fails in terms of its accountability to others. Why, of all the places that a robot could go, would it decide to try and clean the busiest part of the floor? While those in the queue do try to mutually adjust their position for the robot it makes no attempt to acknowledge or accommodate such motion. It wiggles about—unclear in its intentions—ignoring and finally abandoning its attempt to clean around the queue members. Yet the robot also clearly shows a lack of understanding of what those it disturbs are actually doing. As can be seen from the arrangement of the people—at a glance—they are forming a queue \cite{Garfinkel_Livingston_2003}. Our arrangement in space is not just one of movement and efficiency, it can also, at times, have a functional role such as how physically standing somewhere can mean ``I'm first in the queue'' or ``I'm second''. The arrangement of a group in the form of a queue also means ``we're waiting for service'' - giving an 'at a glance' visible reason for why they are standing together. Our position in space has a `purpose', since it makes visible to anyone wanting to join the queue where to go, but it also prevents others from accidentally skipping the queue \cite{livingston_making_1987,brown_order_2004}.

This arrangement provides reasons not to go straight into the middle of the crowd, as this will disrupt the queue. Once actually through there is also then nowhere to actually go—there is just the cafe counter surrounded by people. It is a recurrent feature of public space that groups of people in space who are `together' might be companions, but might also be people arranged in space with a particular purpose in mind. Our example here is a queue—where members of the queue form an order waiting for something together. Queues are a basic organisation of social interaction, and social space, and of course exhibit considerable complexity in both a fundamental way (such as in queuing theory) and in a practical interactional sense (as in studies of different forms of queuing such as cars in traffic, pedestrians in a narrow walkway and so on) \cite{lee_final_1993}. Moving competently in a public space then depends upon not only seeing who is with whom, but also understanding arrangements of people in that space in different groupings. In this case it is `queuing', but other forms might also be `waiting', `moving together', `protesting', `blocking' and so on. That is to say that a basic competence of public space is seeing who is with whom, so as to be able to arrange one's own motion around that.

\citet{turner_notes_1974}'s ``Notes on the Art of Walking'' emphasises the importance of social groups for walking as a pedestrian. These social groups can be groups that are travelling together, who might maintain consistent spatial proximity, a synchronised uniformity of direction and pace, and the management of fleeting physical contact. This might involve engagement in conversation while in motion, mutual orientation such as glances and body alignment. These are crucial indicators that individuals are not just co-present, but intentionally moving as a single, identifiable social entity.

Maintaining conversation while moving as a unit involves considerable management of mutual location, orientation, audibility as well as making available one's gestures to the other. These points have been explored in the robot navigation literature with research exploring how to classify groups automatically \cite{Vazquezetal2015,Nakauchi_Simmons_2002,pmlr-v164-wang22e}, and sociologically in terms of work on `mobile formations'—``[the] practices of forming collectivities (diverse articulations of relations between actors) while on the move'' \cite{mcilvenny_moving_2014}.

As the extract shows, groups can form that are not people who are `together' as in the sense they are connected or even familiar, but rather that they are together functionally—in a queue. Taking into account the functional arrangement of a complex space like an airport thus involves an understanding not only of `who is with whom', in terms of familiarity but also in ad-hoc structures like queues for service.  Indeed, we might even think about how it would be for a robot to join the queue, or at the very least to be able to manage itself so that it does not drive straight into the middle of a queue disrupting the order of service \cite{brown_order_2004}.

\subsection{``What is going on here?'': Scenic intelligibility}
\begin{figure*}[!t]
\centering

\includegraphics[height=0.4\textheight, width=\linewidth, keepaspectratio]{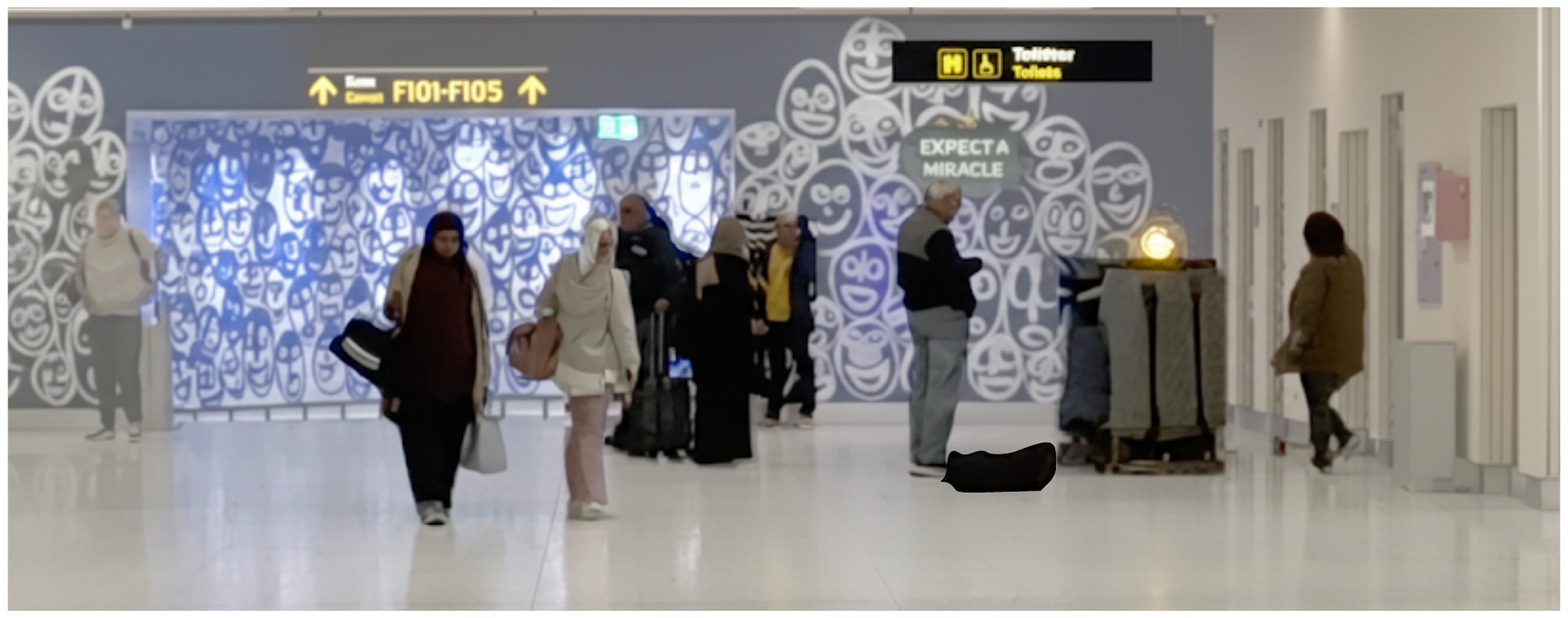}
    \vspace{-5em}
\caption{Robot cleaning in front of people waiting outside the toilet. This image has been enhanced to highlight the waiting passengers and the cleaning robot.}
\label{fig:toilet1}
\Description{}

\vspace{2em}


\includegraphics[height=0.5\textheight, width=\linewidth, keepaspectratio]{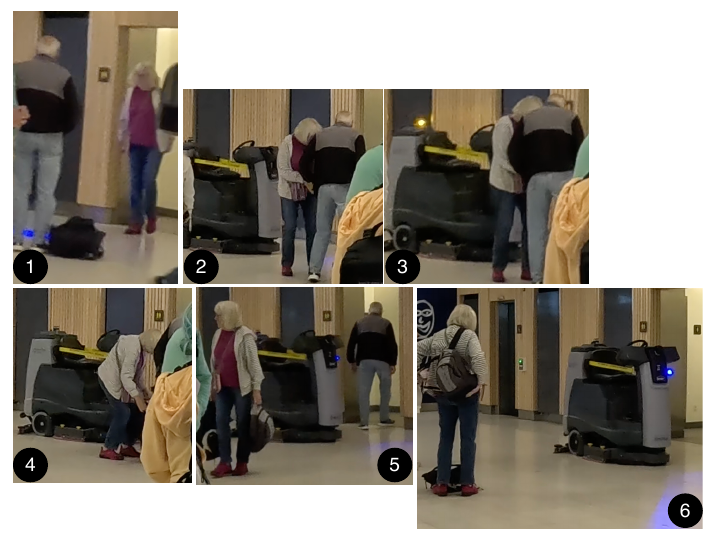}
    \vspace{-1.5em}

\caption{One couple waiting outside the toilet reunite as the woman comes out of the toilet (1), then have to deal with the robot which moves into their baggage (2 \& 3), which they then need to move out of the way (4 \& 5). In the last frame the robot halts outside the toilet door, watched by the woman (6).}
\label{fig:toilet2}

\end{figure*}

Our final extracts (Figures \ref{fig:toilet1} and \ref{fig:toilet2}) show specific activities that are tied to a transit space, which are evident for humans, but not a robot. In the scene captured in \autoref{fig:toilet1}, the robot is cleaning in front of a busy toilet entrance. Eight individuals are either waiting or going into the toilet. In the middle of this, our cleaning robot is moving back and forth, roughly one metre away from the doors to enter the toilets. In the middle of our frame there is a man with white hair waiting for a travel companion who is inside the bathroom, his baggage on the floor at his feet. Behind him there are three groups of passengers with baggage, one woman passing through, and a woman leaving the toilet on the right.

In the short scene depicted in the image, passengers need to make their way around the robot which continues to attempt to clean the airport space as they are waiting. Just a moment later, a man comes out of the toilet, grabs  the suitcase from their travel partner and they both walk off. The man with white hair keeps waiting, guarding his luggage on the floor. There is a rhythmic coming and going, with people entering and leaving the toilets and people waiting for their travel partners to come back. While this is a very mundane scene, and recognisable to anyone who spends time at an airport, the cleaning robot seems unaware of the social organisation of the scene, and unable to adjust its cleaning activity to the rhythms and activities happening in this space. During this ordinary waiting activity, our cleaning robot moves past and manages to clean a bit of the area in front of the toilet as this area is empty for a few seconds. The robot then passes back and forth between the waiting people, with them having to move to accommodate it. 

The challenge here for our robot is twofold. First, there is the overall organisation of this small space—it is a space with travellers walking through quickly. Second, it is a space that travellers pass through to get to and from the entrance to the toilet. It is also a space that is organised by the placement of the toilets—this leads to the natural space for waiting, with passengers standing around their bags. This means that there are multiple obstacles, and a space that will become quiet and busy depending on the flow of passengers over time, and there are groups who are together, and also groups who are together with their baggage. It is into this tight space that the quite large cleaning robot tries to find a way. Alongside not recognising what those in this space are doing, it actually interferes with their movements—coming between passengers and their bags, blocking the entrance to the toilet and so on. It is here that the robot becomes disruptive for a prolonged period, as it attempts to clean close to the door, causing people to need to wait and squeeze around the robot as much as they can.

As humans, one of the skills we use to understand and navigate space is our understanding of the practical purposes of different spaces, and how that comes to influence what we do in that space. We understand that a space has a purpose (e.g. an airport, a toilet, a cafe) and in turn we see the actions of those around in the light of that purpose. But this is a mutual arrangement; often we understand a space through looking at what people are doing there. Therefore, when we find ourselves in unfamiliar spaces we can see that, for example, tea is being drunk and there is a space that looks like a counter—and so we can understand this space as a tea shop. \citet{jayyusi_categorization_2014} refers to this as the ``scenic intelligibility of the social world''. In a social world where signs and the actual demarcation of purposes can often be lightly done or omitted, our ability to reconstruct, to understand different ``scenes'' is essential to everyday competence. Although the design of the airport supports, and makes us expect, different activities in various ways, the design of a space does not in itself determine the activity that may be going on-we can use scenic intelligibility to work out what to do there—perhaps walk around a robot, queue at a cafe or wait outside a toilet. For the people waiting, which could be seen as a problem (why are they waiting here?), what they are doing is made legible by the toilet (they are waiting for the toilet or their travel companions in it). 

Applying \citet{whyte_social_1980}'s work on ``The Social Life of Small Urban Spaces'' to robots, \citet{Pelikan_public-space_2025} remind us of the myriad of activities in public spaces, from street performances to pavement repairs, rhythmic activities during the day, to seasonal and occasional events, events that rely upon recognition amongst passers-by to regulate and produce competent behaviour. This robot does not have these sorts of resources to understand the space, and so moves through the space just like any other space. This leads to ongoing work by everyone else in that space to try and work around its cleaning schedule - moving themselves, moving their bags, squeezing around the robot on their way to the toilet, and so on. 

Expanding the previous case, in  \autoref{fig:toilet2}, we see how two people have to make way for the robot. The travel companion of the man with white hair returns (\autoref{fig:toilet2}.1). The robot comes to clean right in front of them (\autoref{fig:toilet2}.2), and starts to chart a path between them and their bags on the floor (\autoref{fig:toilet2}.3). In airports we are often warned to ``keep our baggage with us at all times'', but this is exacerbated by the bulky nature of that baggage. Clearly bags are `with' people, and separating a group from their bags is something to avoid doing. Yet the robot starts moving towards them and they need to quickly grab their bags and move back (\autoref{fig:toilet2}.4), making space for the robot and retrieving the bag from the robot's path (\autoref{fig:toilet2}.5). The robot then gets stuck again in between all the people coming and going out of the toilet—eventually stopping right in front of the door to the toilet causing an obstruction for those entering and leaving. Those waiting stare at the cleaning robot, unable to affect its movement (\autoref{fig:toilet2}.6).

As we have mentioned the robot again shows its lack of understanding of what to do in this space: the halting motion, the lack of mutual adjustment, failing to understand a queue, failing to see the scenic intelligibility of those waiting for the toilet. But here we can add a final point about how we draw upon our own experiences when we move in a space. For those of us in any space, we are collectively together in that space with a mutual visibility and understanding of each other's actions. This means that there is a reciprocity of perspectives \cite{schutz_collected_1962} between us. Others are experiencing mostly, roughly, the same actions and activities that we are. This means that if we see someone struggling with luggage we can draw upon our own experiences with our own luggage, if we see someone sleeping in a chair, we can understand this from our own experiences of late departures. To understand and interpret others' actions, and so to know how to treat and work with these actions we have our own experiences to draw upon. And it is of course this resource that robots do not have access to—although elements of it can be programmed in (and perhaps simulated), the robot does not need to go to the toilet, or even to keep its heavy bags close by, and this limits its ability to understand the space because it does not act in that space. 


\section{Discussion}
In these examples we have shown how cleaning robots struggle to interact with the inhabitants of the airport, and the movement of passengers in the public transit space. As we outline it here, this demonstrates in different ways how the social aspects of walking in the airport are unavailable to the robots we studied, resulting in some confusions and delays. It may be the case that these robots are not particularly advanced, and it would be unfair to evaluate them in terms of social interaction when clearly they have not been designed to support these interactions. We emphasise, our goal is not evaluation, but instead to use the robots as `breaching devices' in a way to evaluate and understand the arrangement of passenger action in this space \cite{stanley_making_2020}.  

In our discussion, we develop some more general findings for HCI grounded in our empirical observations, exploring in particular how we might design systems that can better adapt to the complexity of transit spaces. We also discuss what role HCI could take in improving robotic systems.  

\subsection{The design space of socially aware movement}
We start our discussion by considering the design space around robot movement. Recent papers in CHI and HRI have explored how robotic movement—in tandem with objects and others—can be designed for, and is an area of considerable complexity that requires further study. For a long time the movements of robots were limited by strong technical limitations, such that the basic kinematic aspects (e.g., staying upright, or avoiding collisions) took priority. Yet recently robots—in particular self-driving cars—have developed complex movements and reactions to others that emulate, at least to some extent, the movements that humans make in those settings. HRI research has started to explore the possibility that the movement of robots need not be seen as something strictly limited for safety reasons, or bounded by the technical capabilities of the architectures used \cite{Hoffman_Ju_2014, cauchard_emotion_2016, sirkin_mechanical_2015}. That is to say, that there might be a \textit{design space} \cite{takayama_putting_2022} opening up where a sensitivity to human interaction and social interaction more broadly becomes the relevant issue \cite{brown2023designing}. In turn this connects with the ongoing discussions in HCI about how the body and technology interact \cite{homewood_tracing_2021}, notably through somaesthetic perspectives \cite{hook_move_2016}. 

Earlier work by \citet{brown_trash_2024} explored how robot trash-cans (albeit human-controlled) could successfully act in a public space with humans. However, in contrast to moving on the city square among people who are seated at tables as described there, the `movement design space' in a transit setting that this paper outlines is perhaps more complex due to its rapidly changing nature. While technical work is of course vital here to actually produce better robots, we would argue that there is also the work of design—for HCI—to scope out what robots should do. This is work that is separate from, albeit eventually dependent upon, technical work to actualise these findings in specific systems. 

Our focus has been on how robots need to be responsive to the social organisation of the space they are travelling through. In our first section we focused on `who goes where', the coordination of trajectories such that (for example) robots and humans can mutually adjust their movements so that both collisions are avoided, but also that `halts', or similar actions that cause disruptions in human movements are minimised. More broadly, this talks to the accountable nature of robots' behaviour \cite{Pelikan_Hofstetter_2023}, that should be designed so that it is understandable to others around. 

We then discussed the need for robots to be able to grasp social groups and `who is with whom', and how understanding the social interactions ongoing between those in the space is also something that needs to be drawn upon. As our example of the queue showed, sometimes collections of people in space can also have a particular meaning (for example, `tour group' or `protest') and in turn this complicates the job of the robot which ideally should not attempt to clean between pedestrians waiting in a queue. Lastly, we examined cases where there are similarly more complex forms of social organisation. Toilets are one frequent case, where robots with simple navigation patterns can end up blocking the door of the toilet as they seek to clean the perimeter of a space.

This leads us to argue for the design space of what we would call `socially aware movement', the design of movement by systems that have both an understanding of space and of what is going on in that space. Humans have an incredible longstanding understanding of situations built upon a lifetime of moving in space with others, and so to carve out individual ‘computationally tractable elements’ is difficult \cite{fischer_progressivity_2019}. Yet, co-present motions will be seen and understood in a social sense by humans, whatever the incompetence or not of an autonomous robot. We would argue then that it is possible to draw upon studies of human-human interaction to understand how system-human motion might unfold, but also to offer directions for how technical work might approach this problem. 

As we have documented even momentary interactions between unacquainted people are social, so how can we trust that these movements are `seen' properly by robots? This would require, as we have outlined above, systems that can mutually adjust their movement with others, such as understanding when a pedestrian turns to avoid the trajectory of the robot, that the robot should in turn adjust itself to maintain the separation. It would also require the ability to understand the basics of groups and those who are together in public space, as well as moving in ways that take into account those groups and configurations in basic ways. In turn, there is the need to understand something of the functioning of the airport, and to respond to different configurations in that airport. More broadly, this is a case of understanding what people are doing in that space, and how different configurations (such as queuing) should lead to different reactions from a robot.

\subsection{How ``good'' is good enough?}
The findings presented in this paper, while not a direct critique of the robots themselves, raise a crucial question for the design and deployment of future robots in public spaces: when it comes to social robot navigation—how good is good enough? While the cleaning robots did relatively poorly in their interactions with those navigating the space, they (mostly) avoided collisions, and any delays and confusions they caused were relatively momentary. For the cleaning staff, the robots save time, as well as helping to present the airport as a modern world-leading airport. In some senses, these robots `were good enough', at least for their current deployment.

But to this we would add two arguments to suggest that these robots are not `good enough'. Public space is by its very nature robust to disruption \cite{stanley_making_2020}. Its open nature means that the order that is established, although fairly minimal, is sufficient to retain a lack of interference of those going about activities in the space. In our case, passengers can pass through mostly unhindered, and staff can do their job, such as keeping a cafe running. We are familiar with disruptions as an everyday part of public space, and have the ability to maintain public order despite those disruptions. It is thus into an already very robust environment that these robots have been introduced. While we have focused on aspects of disruption, it is important to note that the delays and inconveniences that are caused are really only sporadic or short lived. People walk out of the way, or walk around the robot, and continue on their way. It could then be argued that it is of little importance what these robots do, since public space is robust enough to keep functioning. 

Yet even these small disruptions draw a cost amongst passengers, small everyday frustrations as a robot blocks their path or slows them down. What we have in this case is an accumulation of small frustrations. Moreover, while our focus has been on the interactions with those in the airport, behind the scenes the cleaners had to do considerable work to supervise and manage the robots, to diffuse and deal with problems when they arose, and to maintain and position the robots as part of their job. This work—`robot wrangling' \cite{pelikan_people_2025}—has been widely acknowledged in the HRI community, and we would be amiss if we did not mention this work here \cite{takayama_putting_2022}. So while the robots did (mostly) clean, and were considered a success in the airport by those involved, they did not entirely save work for the cleaners, but rather transformed these aspects of their job into maintaining an incompetent robot as part of their own cleaning duties. This means that `good enough', relied upon a large amount of invisible labour by the cleaning staff. Again, it is not that these robots are useless, but rather that they are fragile and partial—tools rather than completely autonomous agents.

Secondly, while our observations show that robots are technically proficient at avoiding collisions, a simple, reactive safety mechanism is insufficient for smooth integration into the social world. For a human, ``good enough'' navigation means more than just not bumping into others; it means moving with a social purpose, understanding the rhythms and flows of a space, and being able to interpret and respond to the actions of others \cite{turner_notes_1974,Hester_Francis_2003}. The robots we studied were ``good enough'' in a technical sense. Their sensors and algorithms were effective at detecting obstacles and triggering an emergency brake to prevent a crash. This level of performance meets a primary, technical requirement: safety. However, we argue that this level of ``good enough'' is flawed. When a robot suddenly halts for no discernible reason to a human observer, it creates friction, confusion, and can even become a new obstacle, as seen in the ``halting problem''. The simple act of avoiding a collision is isolated from the social context in which the interaction is occurring.

The number of robots in public space is still relatively low, but it is possible to foresee a future where robots are commonplace. Traffic jams of different sorts with delivery robots in US cities have already been reported, and so it could well be the case that these emergent problems will grow in magnitude as the deployments grow. If we think about the possibility of social environments—including but going beyond the airport as a transit space—becoming more busy with robots, we have to face that their incompetent movements would go from being an inconvenience to blocking and disrupting the ordinary activity of these places. The goal should be a robot that is not just safe, but also predictable and legible to the humans around it. This requires moving from a robot-centric view—where the robot’s primary task is to get from A to B while avoiding collisions—to a human-centric view, where the robot's movement contributes to the overall social order of the space it inhabits.

\subsection{Strong concepts for robot movement}
Our final discussion item considers the conceptual implications of our findings, arguing for the need to move to a more socially rich notion of robot movement. Our goal has been to enrich the understanding of robot movement as a problem of path planning and obstacle avoidance that dominates research on \textit{robot navigation} \cite{Francis-socialnavigation2025, mavrogiannis_core_2023} with an HCI perspective. Specifically, we have explored how ideas from the sociology of walking could be integrated into a conceptual framework for robot movement. Crucially, this means shifting from a focus on maintaining a safe distance and interacting with individual humans, to a relational perspective that incorporates the social fabric of public spaces. Movement has been repeatedly demonstrated to be not a solitary act, but an interactional and collaborative accomplishment \cite{deppermann_overtaking_2018, broth_walking_2013, turner_notes_1974}. A robot’s movement is not just about its trajectory but how that trajectory is interpreted and responded to by others \cite{Pelikan_2021}. For example, the ``halting problem'' shows that a robot's lack of legibility—its inability to signal its intent—disrupts the fluid, give-and-take nature of human movement.

One way of exploring this further could be to develop ``strong concepts'' for designing socially competent robots \cite{hook_strong_2012}. Strong concepts are ideas that promote a higher level understanding of a design problem or interaction. They are generative for design by not just describing single situations, but offering insight into problems and new design possibilities, contributing a form of ``intermediate-level knowledge'' that sits in between specific examples and high-level theories \cite{Lupetti_Zaga_Cila_2021}. Drawing from our analysis, we suggest three potential strong concepts that could inform the design of more socially competent robots:

\begin{description}
        
\item[Sequentiality of Movement] Human coordinated movement is not a series of isolated actions, but reorientations and adjustments that are sequentially and micro-sequentially \cite{mondada_collaboratively_2022} organised. People \textit{recognize} and \textit{anticipate} how others move, reading the ``in order to do X'' orientation. Motion design for robots in public needs to consider how robots can contribute to recognisable, shared sequences of actions. What are people doing and where will they move next? What is the robot doing and where is it going to move next? For example, a robot at an elevator should position itself to signal that it is waiting, rather than simply stopping close by.

\item[Membership Categorization] Competent human navigation relies on recognising who people are in a space—a family, a group of business travellers, or a service worker \cite{Hester_Francis_2003}. This understanding shapes how we move and has implications for norms of who should give way to whom. A robot that moves in public space should have the ability to categorise the ``units'' of people it is navigating around, e.g. groups and queues but it also needs some basic way to distinguish different kinds of people and to identify people with specific roles \cite{pelikan_encountering_2024, Reeves_Pelikan_Cantarutti_2025}. This would include being able to read the social fabric, who these people are and what they might be up to, what kind of movement we expect of them, and what actions are tied to these categories. In the social world, such phenomena are ``inference rich'', allowing members to anticipate and expect certain actions and not others. A cleaning robot that could tell the difference between a person waiting in a queue and someone standing idly outside a door with an excessive amount of luggage would have a better basis on which to plan its cleaning trajectory.

\item[Scenic Intelligibility]
Above we discussed how, at a glance through our understandings of particular spaces, we can understand what is expected to be done there—the ``scenic intelligibility'' \cite{jayyusi_categorization_2014}. This means that to understand a particular space, a robot needs to understand what goes on there, and then use that as a resource for understanding how to regulate its movement in that space. Although the design of the airport supports, and makes us expect, different activities in various ways, the design of a space does not in itself determine the activity that may be going on \cite{mcilvenny_communicating_2009}. Robot design needs to acknowledge different types of activity rhythms, some of which may be easier and some harder to anticipate \cite{pelikan_making_2025}. For instance, at some times during the day, a particular area in the airport will be entirely empty and at other times crowded. A cleaning robot that could tell the difference would be able to organise its job more efficiently.

\end{description}

Moving toward these strong concepts for robot movement requires a fundamental shift in design philosophy. Instead of asking, ``How can the robot safely get from A to B?'', we should be asking, ``How can the robot participate in the social world of getting from A to B?'' This moves the focus from a purely technical challenge to a socio-technical one, where successful robot design is measured not just by collision avoidance but by its seamless, invisible, and unproblematic integration into the everyday rhythms of public life.

Through the strong concepts, we hope to provide designers with a starting point for what to design for in social navigation. The translation of the strong concepts into engineering practice is beyond the scope of this paper, and requires careful exploration and in situ testing. We hope to outline here the types of challenges that social robot navigation research should set out to address, exploring for instance how planning algorithms and technical systems can best support the mutual coordination of movement between robots and humans. Our work also outlines current challenges for robot sensing, providing detailed examples of the kinds of things robots may need to detect in human spaces. 

\section{Limitations}
Before we conclude, we should outline some of the limitations in this work. Clearly it builds on a relatively small corpus of video data, with only two robots in one airport. As such, it rests on analytic generalisability rather than quantitative analysis, and the examples we provide can be seen as a detailed analysis of revealing edge cases. Our results are tools to think through robot behaviour and of course do not translate to all public contexts. Airports are distinctive settings, and studies in other transit settings may yield additional insights. Our fieldwork was relatively short in duration and long-term investigation could have resulted in more deviant observations. As researchers, we are always also participants in the settings we study. By visibly recording the robots they may have received more attention by travellers than on other days. 

\section{Conclusion}
Our analysis of cleaning robots in a public airport highlights that while these robots may be technically capable of avoiding collisions, their navigation lacks a crucial social understanding of human movement, group dynamics, and the situational context of action in a transit space. We argue that a purely technical approach fails to account for how humans collaboratively organise their movement in activity-bound places. The problems we observed, from the ``halting problem'' to the robot's lack of understanding of social formations and the purpose of specific spaces, are not minor technical glitches but fundamental breakdowns in the robot's ability to participate in a social world. This suggests that for robots to be ``good enough'' for public deployment, their design must move beyond simple, reactive safety mechanisms to embrace a more nuanced, human-centric understanding of movement as an interactional, collaborative accomplishment. This shift would recognise that successful robot integration depends not just on technical performance but on the robot's ability to be a legible, predictable, and socially aware participant in the everyday rhythms of public life. This needs robots that reduce - and do not just reconfigure - the labour required of human staff need take care of the robots in their daily work.

\begin{acks}
We would like to acknowledge WASP-HS ("AI in Motion: Studying the Social World of Autonomous Vehicles", MMW 2020.0086), and Villum Fonden ("Hybrid senses", 69162) for financial support for this project.  We also thank our airport collaborators for being generous with their time and help with this project.
\end{acks}

\balance

\balance


\bibliographystyle{ACM-Reference-Format}
\bibliography{references,additional-refs}

\end{document}